\documentclass[a4paper,fleqn]{cas-sc}

\usepackage[numbers]{natbib}
\pdfoutput=1

\def\tsc#1{\csdef{#1}{\textsc{\lowercase{#1}}\xspace}}
\tsc{WGM}
\tsc{QE}

\begin{document}
\let\WriteBookmarks\relax
\def\floatpagepagefraction{1}
\def\textpagefraction{.001}

\shorttitle{Defects drive the tribocharging strength of PTFE}

\shortauthors{A. Ciniero, G. Fatti, M. Marsili, D. Dini, and M.C. Righi}

\title [mode = title]{Defects drive the tribocharging strength of PTFE}                      
\tnotemark[1,2]



%
\author[1]{Alessandra Ciniero}
\credit{Conceptualization; Investigation; Visualisation; Formal analysis; Writing - review editing; Funding acquisition.}

\affiliation[1]{organization={Department of Mechanical Engineering, Imperial College London},
    addressline={South Kensington Campus}, 
    city={London},
    postcode={SW7 2AZ}, 
    country={UK}}

\author[2]{Giulio Fatti}[orcid=0000-0001-7000-7310]
\credit{ Investigation; Visualisation; Formal analysis; Writing - review editing. }
\author[2]{Margherita Marsili}[orcid=0000-0003-1009-287X]
\credit{Formal analysis; Visualization; Writing - original draft.}

\affiliation[2]{organization={Department of Physics and Astronomy, University of Bologna,},
    addressline={Viale Carlo Berti Pichat 6/2}, 
    city={Bologna},
    postcode={40127}, 
    country={Italy}}

\author[1]{Daniele Dini}[orcid=0000-0002-5518-499X]
\credit{ Formal analysis; Writing - review editing; Funding acquisition.}

\author[2,1]{M. Clelia Righi}[orcid=0000-0001-5115-5801]
\credit{Conceptualization; Formal analysis; Supervision; Writing - review editing; Funding acquisition.}
\cormark[1]
\ead{clelia.righi@unibo.it}
\ead[URL]{www.tribchem.it}

\cortext[cor1]{Corresponding author}

\begin{abstract}
If polytetrafluoroethylene (PTFE), commonly known as Teflon, is put into contact and rubbed against another material, almost surely it will be more effective than its counterpart in collecting negative charges. This simple, basic property is captured by the so called triboelectric series, where PTFE ranks extremely high, and that qualitatively orders materials in terms of their ability to electrostatically charge upon contact and rubbing. However, while classifying materials, the series does not provide an explanation of their triboelectric strength, besides a loose correlation with the workfunction. Indeed, despite being an extremely familiar process, known from centuries, tribocharging is still elusive and not fully understood. \\
In this work we employ density functional theory to look for the origin of PTFE tribocharging strength. We study how charge transfers when pristine or defective PTFE is put in contact with different clean and oxidised metals. Our results show the important role played by defects in enhancing charge transfer.
Interestingly and unexpectedly our results show that negatively charged chains are more stable than neutral ones, if slightly bent.
Indeed deformations can be easily promoted in polymers as PTFE, especially in tribological contacts. These results suggest that, in designing materials in view of their triboelectric properties, the characteristics of their defects could be a performance determining factor.
\end{abstract}

 \begin{graphicalabstract}
\includegraphics[width=0.5\textwidth]{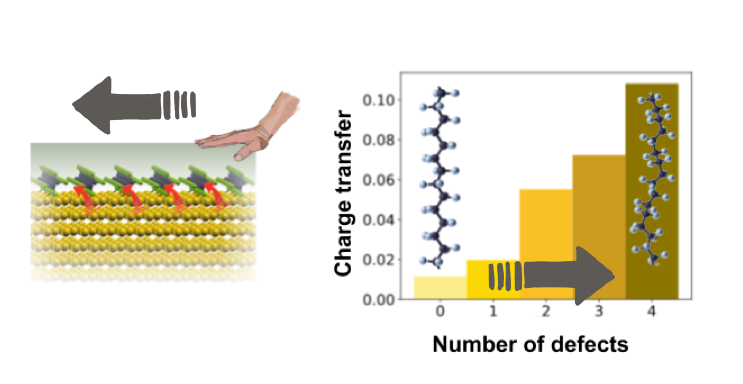}
 \end{graphicalabstract}

\begin{highlights}
\item Electron transfer between pristine and defective PTFE chains and different clean and oxidised metals has been quantified using DFT.
\item The presence of defects on PTFE chains strongly enhances the charge transfer. 
\item All the analysed defect configurations are stabilized by negative charging. Negatively charged bent chains, are more stable than neutral unbent ones. 
\end{highlights}

\begin{keywords}
triboelectric effect \sep charge transfer \sep density-functional theory \sep PTFE \sep defects
\end{keywords}

\maketitle

\section{Introduction}

\begin{figure}[htb!]
\centering
\includegraphics[width=0.5\textwidth]{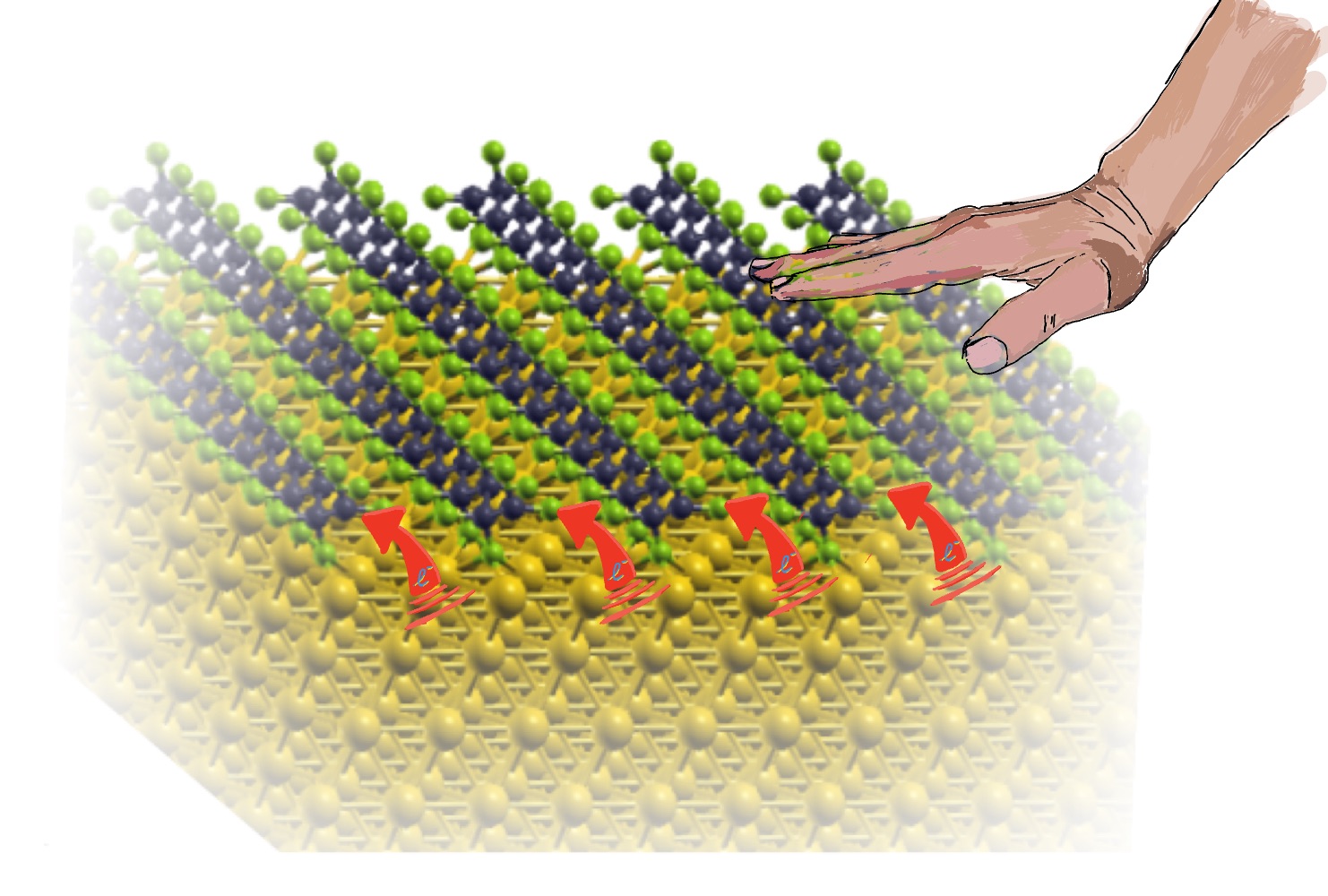}
\caption{\label{fig:geometry} Cartoon of the PTFE/Au system and of the tribocharging process. The PTFE chains, in the high pressure phase, are set parallel to the metal surface. Pictorially represented as the red arrows, electrons are transferred towards the PTFE chains upon contact and rubbing. PTFE is one of the most efficient material in acquiring electrons through tribocharging.}
\end{figure}

Tribocharging, the electrostatic charging of two materials upon contact and rubbing, is the basic phenomenon by which most of us have been introduced to the study of electricity. It is an ordinary process present in our daily life, sometimes with unpleasant results as we touch our car's door handle! Though familiar and well known from centuries, the triboelectric effect is subtle and still not fully understood: for instance the fundamental and possibly distinct roles of contact and rubbing are still to be clarified: it is not clear even if electrons, ions or bulk material are transferred, and predictions on the direction and amount of charge flow are made on the basis of an empirical triboelectric series, ranking materials according to their aptness at losing or gaining electrons \cite{Lacks_nat_chem_rev2019}. Only recently, and for a specific set of polymers and non-metallic materials, is this series being coherently quantified with measures of triboelectric charge density under controlled and standardized conditions \cite{Zou_nat_comm2019,Zou_natcom2020}. Controlling and engineering the triboelectric effect would be important, not only to avoid hazardous static charging but also for the development of triboelectric nanogenerators, novel devices that harvest mechanical energy, that would otherwie be lost by converting it into electricity \cite{Kim2020,Liu_natcomm_2022}.

Polytetrafluoroethylene (PTFE), a synthetic fluoropolymer, commonly known with its commercial name Teflon, ranks extremely high in the tribolectric series.
Notably, in Ref. \cite{Zou_natcom2020}, PTFE has been chosen as a reference material.
Despite the widespread use of this polymer, it is actually very difficult to link the triboelectric strength of PTFE to an intrinsic property of the material itself.
The remarkable electron attraction toward PTFE is indeed unexpected as this strongly insulating material is exposing negatively charged
fluorine atoms at the exterior of the carbon chains.
On the contrary, polyethylene (PE), the hydrocarbon analogue of PTFE,
while still ranking very high in the tribolectric series,
is less effective than PTFE in tribocharging despite its partially positively charged hydrogen terminations and despite the presence of surface trap states, absent in
the band structure of PTFE high-pressure phase surfaces \cite{righi_PhysRevLett2001,fatti_acsappl_pol_mat2020}.
Unveiling what makes PTFE so effective in collecting electrons even with its apparent unfavorable properties could allow the implementation of alike mechanisms and conditions to reduce or enhance tribocharging also in other systems.

A common feature between PTFE and PE is the possibility of easily presenting structural defects:
PTFE (and PE) chains are typically bent,
and within tribological contact processes their structures can further deform or break.
Indeed in Ref. \cite{fatti_acsappl_pol_mat2020}, we showed that fluorine vacancies in PTFE
high-pressure phase surfaces were able to generate trap states in the band structure of the material.
Here we go beyond this work using density functional theory (DFT) calculations to directly and fully simulate how electrons transfer when pristine or defective PTFE is put in contact with different metals or metal oxides.
Our study reveals the importance of defects in enhancing the process of tribocharging: indeed, regardless of the material in contact with PTFE, charge transfer strongly increases in the presence of defective chains. We ascribe such charge transfer enhancement to the defective PTFE property of becoming more stable by negative charging.
Interstingly, and unexpectedly, we found that defects originating from chain deformations are more stable, upon negatively charging, than
those involving bond breaking.

These results suggest a general mechanism:  while the direction and amount of charge transfer upon friction-less, ideal static contact between pristine materials is mainly dictated by band structure and relative electronic energy level distribution \cite{Zou_natcom2020}, in the presence of realistic contact pressures and/or rubbing it is also influenced by the defects that may be created, which, if of the proper kind, further contribute to the material charging.

\section{Methods}
All calculations were performed employing DFT as implemented in the Quantum ESPRESSO package \cite{Giannozzi_2017}. 
Ultrasoft pseudopotentials, the PBE  \cite{perdew_prl1996} exchange–correlation functional and semiempirical Grimme's DFT-D2 Van der Waals correction \cite{grimme_JCC2006} were employed throughout the calculations. Atomic position were relaxed until all force components were smaller than 0.001 a.u. 

As commonly performed in plane-wave DFT calculations, imposing periodic boundary conditions to the system, the metal(metal oxides)/PTFE interfaces are modeled by creating supercells that, looking along the z-direction (perpendicular to the surfaces), comprise a PTFE-covered metal(metal oxides) slab and a vacuum layer, deep enough to avoid spurious interactions between the replica. In the xy plane, the PTFE chains develop along the y direction and the metal (metal oxide) unit cell is repeated to accomodate the periodicity of the PTFE monolayer. Slab thicknesses were converged with respect to surface energy. All relevant computational parameters are summarised in Tab. \ref{tab:comp_param}.

\begin{table}[htb!]
    \centering
    \begin{tabular}{|c|c|c|c|c|c|c|}
    \hline
     Surface    & $E_{cut}$& N$_{l}$& N$_{at}$& N$_{PTFE}$ & Vacuum & K-point      \\
                & (Ry)       &        &         &            & (\AA{})  & mesh         \\
    \hline
     Fe(110)    &  240     & 4    & 24    & 12       & 22     & $4\times2\times1$      \\
     Cu(111)    &  240     & 4    & 16    & 8        & 27     & $1\times1\times1$      \\
     Au(111)    &  280     & 6    & 16    & 8        & 16     & $3\times3\times1$      \\
     FeO(110)   &  280     & 4    & 36 (9)& 12       & 27     & $1\times1\times1$      \\
    Cu$_2$O(111)&  280     & 5    & 24    & 16       & 12     & $3\times3\times1$      \\
     \hline
    \end{tabular}

    \caption{Relevant computational parameters. $E_{cut}$ is the kinetic energy cutoff for the plane wave expansion of the density. N$_{l}$, N$_{at}$, and N$_{PTFE}$ are the number of atomic layers that make up the metal (metal oxide) slab, the number of atoms within each atomic layer, and the number of PTFE monomer in each simulation unit cell. For the FeO(110) surface the number in parenthesis represents the number of oxygen adatoms adsorbed on the iron surface. }
    \label{tab:comp_param}
\end{table}

\section{Results and discussion}
We studied the charge transfer between PTFE and representative metals (Au, Cu, Fe) and oxidized metal (Cu$_2$O, Fe:O) substrates. 
PTFE chains in form III, i.e. the stable phase in compressed conditions, as those present in tribological contacts \cite{fatti_JPCC2019}, are positioned parallel to the metal substrates. In this way a more stable interface is formed with respect to the case of perpendicular chains \cite{WU_nanoenergy2018}. As an example, in Fig. \ref{fig:geometry}, the geometry of PTFE deposited on the Au surface is pictorially shown.

As shown in Tab. \ref{tab:prop_pristine_surface}, PTFE binds significantly stronger to iron with respect to copper and gold, due the different reactivity of the three surfaces.  The presence of oxygen, passivating the surface, reduces drastically the adhesion strength of PTFE from -0.53 J/m$^2$ to -0.18 J/m$^2$. At the same time the workfunction of the oxygenated surface increases, in line with Ref. \cite{fatti_jpcc2018}. On the contrary, the presence of under-coordinated copper atoms in the case of copper oxide strongly favours PTFE adhesion, similarly to what found concerning graphene adsorption in Ref. \cite{antonov_nanoscale_adv2022}.

\begin{table}[htb!]
    \centering
    \begin{tabular}{|c|c|c|c|}
    \hline
     Surface        & Surface energy& Work function & Adhesion  \\
                    & (J/m$^2$)     & (eV)          &  (J/m$^2$)\\
    \hline
     Fe(110)        &  2.5          & 4.9           & -0.53     \\
     Cu(111)        &  1.3          & 5.2           & -0.16     \\
     Au(111)        &  0.7          & 5.1           & -0.14     \\
     FeO(110)       &               & 5.4           & -0.18     \\
    Cu$_2$O(111)& 0.8 \cite{soon_PhysRevB2007}& 4.8 & -0.53     \\
     \hline
    \end{tabular}

    \caption{Surface energies and work function of the clean surfaces. Adhesion energies and
 are computed with respect to pristine PTFE. FeO surface energy is not present because no bulk reference value is available since oxidation is simulated through the presence of O adatoms.}
    \label{tab:prop_pristine_surface}
\end{table}

\begin{figure}[htb!]
\centering
\includegraphics[width=0.6\textwidth]{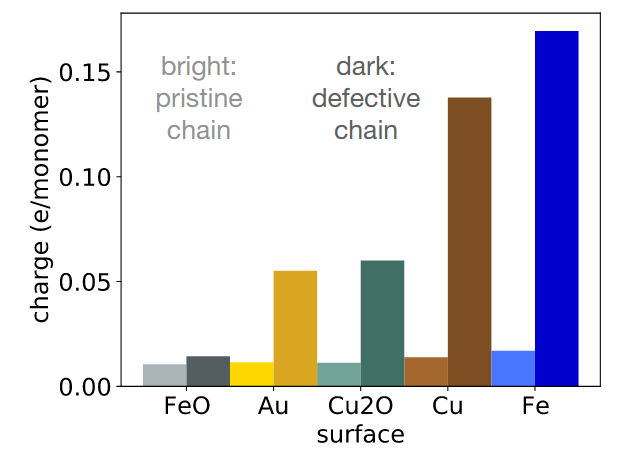}
\caption{\label{fig:charge_transfer} Electron transfer per PTFE monomer. Lighter colours refer to pristine chains, darker color to defective, 12.5\% defluorinated chains.}
\end{figure}

Confirming the high ranking of PTFE in the triboelectric series, all substrates transfer between 0.01 and 0.02 electrons per monomer to pristine PTFE, shown in brighter colours in Fig. \ref{fig:charge_transfer}. The direction of charge transfer is justified by the relative alignment of the work function: PTFE workfunction (5.7 eV -5.8 eV \cite{Trigwell_IEEE2003,Kazuhiko_phys+scr_1990}) is larger than the work function of all chosen substrates. Despite the small numbers involved, we can recognize that in the case of the clean metal surfaces larger adhesion energies correspond to larger charge transfers. Indeed it has been shown that adhesion energy has a strong correlation with charge accumulation at the interface \cite{Wolloch-18,reguzzoni2012} 
The Cu$_2$O surface does not seem to follow this general trend, most probably due to the local interactions, connected to the presence of undercoordinated oxygen atoms, that play a major role in this system.

Interestingly, despite the overall negative charge flow towards the adsorbate, the work function difference, $\Delta \phi=\phi-\phi_0$, is negative for all the substrates. Here $\phi$ and $\phi_0$ are the covered and clean surface's work functions respectively. This finding is only apparently counter intuitive as a direct proportionality between the overall charge transfer and work functions is found only in the most simple cases, such as adsorption of selected atomic species \cite{arefi_JPCC2014,marri_PCCP2020}. In general situations, work function changes are the result of the complex charge redistribution over the entire surface region and can be negative also in the presence of a net electron flow from the substrate to the adsorbate \cite{Michaelides_PRL2003,fatti_jpcc2018}.

\begin{figure}[htb!]
\centering
\includegraphics[width=0.6\textwidth]{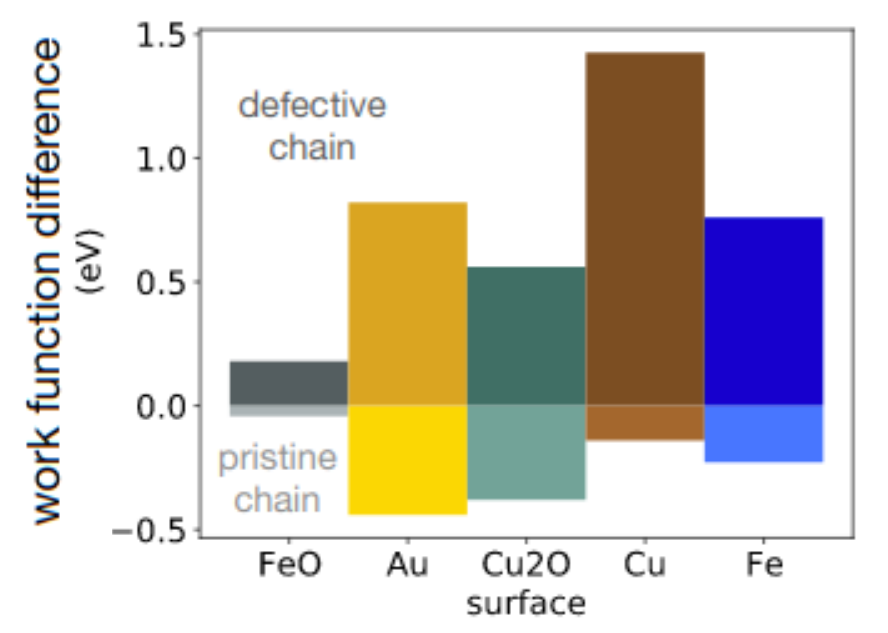}
\caption{\label{fig:wf_change} Work function change between clean and covered surface. Bottom bars refer to the pristine chain, top bars refer to defluorinated chains.}
\end{figure}

At the nanoscale, in tribological conditions, material experience extremely high mechanical forces and thermal vibrations \cite{BOGDANOVICH_tribology_international2006}. In the case of metal/PTFE friction, such high contact stresses and hot spots are able to provide enough energy to drive complex tribochemical reactions and even break PTFE chains\cite{onodera_JPCC2014,harris_macromolecules_2015}. Chain defluorination is one of the possible processes induced by frictional contact: indeed the presence of metal fluorides is often detected in tribological experiments concerning metal/PTFE interfaces and moreover chain deflourination has been theoretically found as one of the first process happening upon sliding contact \cite{onodera_JPCC2014,jintang_wear2000,ZUO_wear_2014}. This is important because chain defluorination has strong impact on the electronic properties of PTFE leading to the formation of surface states within the fundamental gap \cite{fatti_acsappl_pol_mat2020}. Such surface states, that act as deep traps for electrons, would be otherwise absent in the case of pristine chain, and are instead found in pristine polyethylene, another extremely high ranking material in the tribological series \cite{righi_PhysRevLett2001}.

Defluorinated chains were thus put in contact with our substrates. The charge transfer towards 12.5\% deflourinated chains, compared to the pristine chain cases, is shown in Fig. \ref{fig:charge_transfer}. For all substrates we report an increase in charge transfer, which in the case of pure copper is more than tenfold. At the same time, as shown in Fig. \ref{fig:wf_change}, the workfunction change upon defective chain adsorption turns positive for all cases as a consequence of the increased magnitude of electron transfer towards the adlayer.

\begin{figure}[htb!]
\centering
\includegraphics[width=0.8\textwidth]{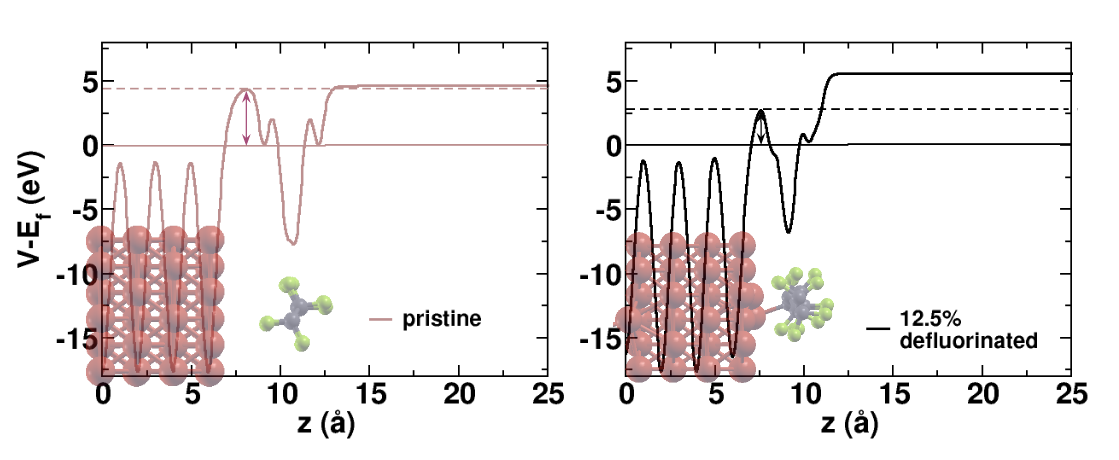}
\caption{\label{fig:interface_barrier} Planar average of the electrostatic potential of PTFE on Fe. Left panel: pristine chain. Right panel defective, 12.5\% defluorinated chain. Energies are referred to the Fermi level, the arrows indicate the interface potential barrier.}
\end{figure}

The charge transfer enhancement in the presence of defected chains is associated the lowering of the interface potential barrier that electrons must overcome in order to reach the adsorbate. In Fig. \ref{fig:interface_barrier} the iron/PTFE electrostatic potential planar averages, featuring the lowering of interface barrier, are shown.

\begin{figure}[htb!]
\centering
\includegraphics[width=0.6\textwidth]{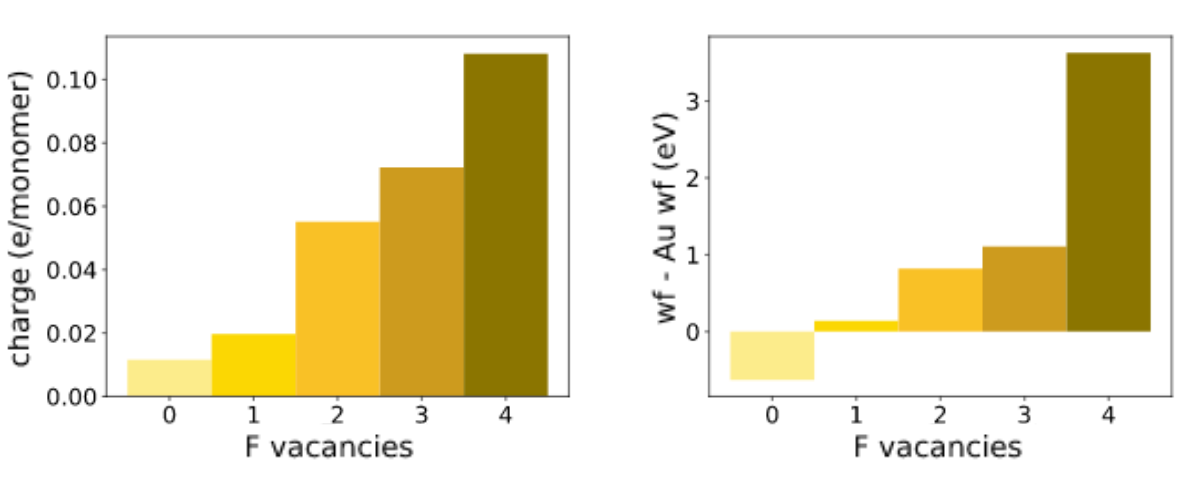}
\caption{\label{fig:charge_transfer_vs_ndefects} Left panel: electron transfer per chain monomer for the PTFE/Au interface increasing the number of F vacancies in the chain. Right: workfunction variation between PTFE/Au and clean gold surface upon increasing the number of F vacancies.}
\end{figure}

By increasing the density of defects the amount of charge transfer increases more than linearly, as can be observed in the left panel of Fig. \ref{fig:charge_transfer_vs_ndefects} for the case of Au, meaning that the interaction between defects even further favours electron transfer. The work function difference closely follows the trend of charge transfer, as shown in the right panel of Fig. \ref{fig:charge_transfer_vs_ndefects}.

The great tendency of defluorinated chains to acquire electrons is strictly connected to the fact that the defective, negatively charged chains are more energetically stable than neutral ones \cite{fatti_acsappl_pol_mat2020}. This, on the contrary, does not happen in the case of pristine chains.

\begin{figure*}[htb!]
\centering
\includegraphics[width=\textwidth]{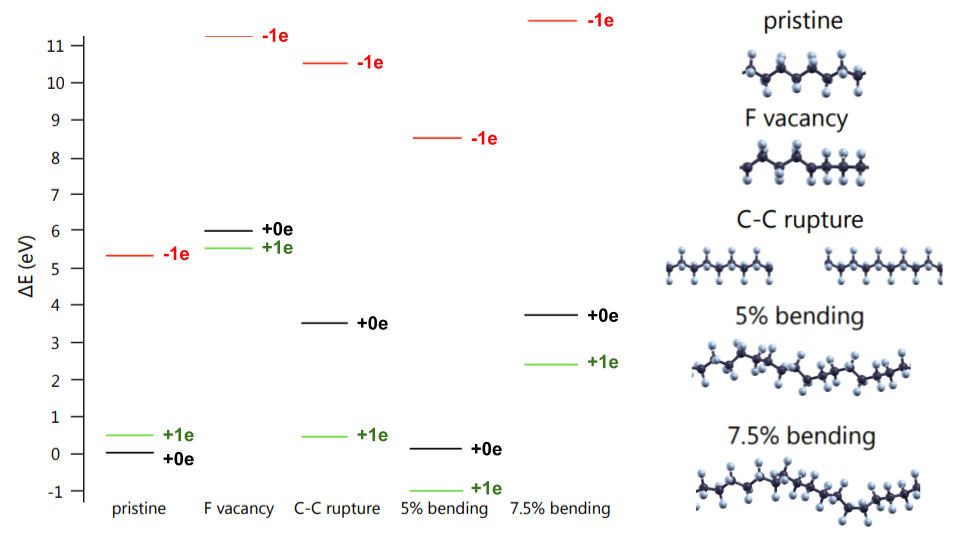}
\caption{\label{fig:defects} Left panel: total energies of neutral and charged pristine and defective PTFE chains with respect to the total energy of the pristine neutral chain. Right panel: different geometries of the investigated systems. }
\end{figure*}

Besides defluorination, the processes and defects that can be possibly induced by contact and rubbing may be of many different kinds: in particular chains may bend and C-C bonds may break \cite{onodera_JPCC2014,jintang_wear2000,ZUO_wear_2014}. If also these defects present an electronic behaviour similar to that of fluorine vacancies, namely if they are also stabilized by negative charging, this could link the triboelectric strength of PTFE to the characteristcs of its defects more than to some intrinsic property of its pristine form. For this reason, we studied the electronic properties of PTFE chains hosting either a C-C rupture, or a 5\% and 7.5\% bending, the geometries of these systems are shown in the right panel of Fig. \ref{fig:defects}.

From the diagram in the left panel of Fig. \ref{fig:defects}, it is clearly seen that pristine chains favour being neutral rather than negatively or positively charged. At the same time, as neutral structure it is the most stable, confirming its ground state nature. On the other hand, all the defects that we considered here are stabilized by negative charging. Interestingly, the negatively charged, 5\% bent chain has a negative formation energy with respect to the pristine neutral structure. Indeed, for polymers, where deformation and bending can be easily expected in particular upon frictional contact, the strong stabilization that the mildly bent chains present is very significant in view of tribocharging.

From an electronic structure perspective, also for these defects, the stabilization upon charging may be ascribed to the population of trap states within the gap. Indeed, whereas in the case F vacancies a single electronic state is present in the gap, C-C rupture provides two states in the gap. Interestingly, chain deformations have also been linked to an electronic structure more favourable for tribocharging \cite{Kang_nanoenergy2022}, and  an interplay between the triboelectric and flexoelectric effects has been shown \cite{mizzi_prl2019}.    

\section{Conclusions}
In conclusion the triboelectric effect, well known from centuries, always remained elusive and hardly understood and predictable, if not for phenomenological and qualitative series ranking materials with respect to their ability of positively or negatively charging when put into contact. Besides a loose relation with work function for metals\cite{Zou_natcom2020}, no intrinsic property of materials has been clearly linked to triboelectric strength. For instance in the tribolectric series polyethylene, the hydrocarbon counterpart of PTFE, ranks lower than PTFE. This happens despite the presence of partially negatively charged fluorine atoms exposed by PTFE.\\
In this work we have shown how, in addition to intrinsic properties of pristine materials, also the characteristics of defects that frictional contact may induce in them must be taken into account to fully quantify its charge collection capability. Indeed at the nanoscale, during frictional processes, high temperatures and stresses might be generated driving the formation of a large variety of defects. In particular, we have shown how defluorinated chains allow a substantial gain of electron transfer towards PTFE for all the substrates under investigation. This property is linked to the energetic stabilization of this type of defect by negative charging. Also the other defects that we investigated (C-C rupture and bending at various degree) share the same feature, suggesting that the origin of the triboelectric strength of PTFE lies in the electronic properties of its defects. Especially interesting is the case of mildly bent chains whose formation energy, in its charged form, is negative with respect to pristine neutral chains, indicating that negatively charged chains are more stable than the neutral ones, if slightly bent. And indeed, deformation and bending can easily happen in tribological conditions of polymers such as PTFE. Finally, besides fundamental understanding, these results suggest a general mechanism that could be possibly exploited to enhance or reduce tribolectric charge transfer by controlling the properties of material's defects.

\appendix
\section{Acknowledgments}
This work is part of the "Advancing Solid Interface and Lubricants by First Principles Material Design (SLIDE)" project that has received funding from the European Research Council (ERC) under the European Union’s Horizon 2020 research and innovation program (Grant agreement No. 865633). DD acknowledges the support received by the Engineering and Physical Sciences Research Council (EPSRC) via his Established Career Fellowship (EP/N025954/1), and the Royal Academy of Engineering for the Research Chair in Complex Engineering Interfaces. AC thanks the European Union Horizon 2020 Marie Curie Actions for supporting this work under the Grant Agreement No.798245. MM acknowledges the CRESCO/ENEAGRID High Performance Computing infrastructure and its staff \cite{cresco}
CRESCO/ENEAGRID High Performance Computing infrastructure is funded by ENEA, the Italian National Agency for New Technologies, Energy and Sustainable Economic Development and by Italian and European research programmes, see
http://www.cresco.enea.it/english for information. MM would like to thank Dr. Paolo Restuccia for fruitful discussions.

\printcredits

\bibliographystyle{cas-model2-names}
\bibliography{biblio}

\newpage

\bio{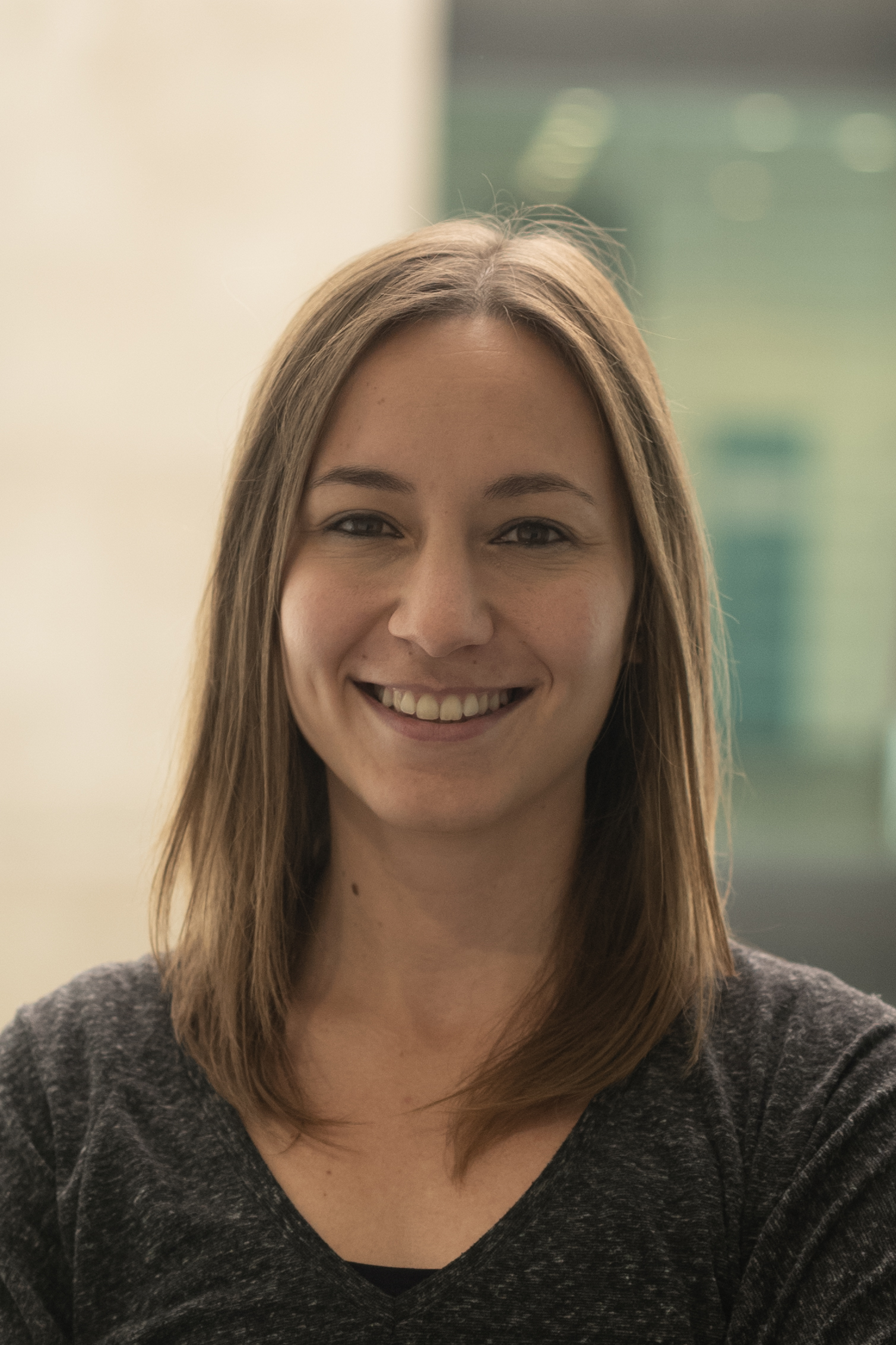}
Alessandra Ciniero obtained her PhD in Mechanical Engineering at the Imperial College of London (UK). Afterwards, she worked as a research fellow after being awarded an EPRCS Doctoral Prize Fellowship and a Marie Curie Individual Fellowship. Her research focused on the investigation of the tribocharging phenomenon with the aim of redesigning materials used for triboelectric nanogenerators to increase their energy storage/supply efficiency. She is currently an innovation consultant focusing on venture building and advanced materials commercialisation.
\endbio

\bio{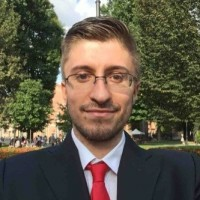}
Giulio Fatti is currently a postdoctoral researcher at the Korea Institute of Ceramic and Technology. He has received his Ph.D. at the University of Modena and Reggio Emilia, Italy, under the supervision of Professor M.C. Righi. His research interests span from tribology to energy materials. His current research focus is the investigation of the triboelectric effect, aiming at unveiling its underlying mechanism to provide guidance for the development of triboelectric nanogenerators.
\endbio

\bio{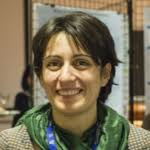}
Margherita Marsili is Assistant Professor at the Physics and Astronomy Department of the University of Bologna (Italy). Within her research activity she has been carrying out  advanced first-principle simulations of the electronic and optical properties of materials including methodological developments and implementations in open-source publicly distributed codes (under GPL licence) such as Quantum Espresso, and  WaveT-TDPLAS and Yambo. She has recently joined the Materials Modeling \& Tribology group at the University of Bologna, focusing her research on the study of solid-solid interfaces.
\endbio

\bio{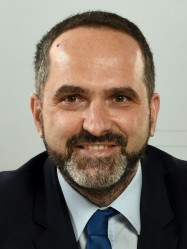}
Daniele Dini is Professor in Tribology at Imperial College in London and Head of the Imperial College Tribology Group. Professor Dini’s research centres on the application of advanced modelling strategies to applied mechanics, materials, physics, chemistry, biomechanics and structural integrity, with a particular focus on tribology. Professor Dini is author/coauthor of more than 250 scientific publications and the recipient of a number of awards, including the Tribology Trust Bronze and Silver Medal awarded by the IMechE (2004 and 2022 respectively) and the inaugural Peter Jost Tribology Award (2021), received from the International Tribology Council. He is a Fellow of the Royal Academy of Engineering, the Institute of Physics, the Institution of Mechanical Engineers and the Society of Tribologists and Lubrication Engineers.
\endbio

\bio{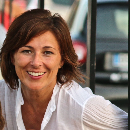}
M. Clelia Righi is Full Professor at the Department of Physics and Astronomy of Bologna University, leading the Materials Modeling \& Tribology group. Her research activity focuses on the development and application of computational methods to understand and predict the behaviour of materials from first principles, particularly of surface and interface phenomena. She adopted pioneering computational approaches in tribology and applied them for understanding chemical reactions activated by mechanical stresses and designing materials to reduce friction. In 2019 she received an ERC consolidator grant for the project “Advancing solid interfaces and lubricants by first principles material design” (SLIDE).
\endbio

\end{document}